\documentclass[aps,twocolumn,pre,showpacs,preprintnumbers,amsmath,amssymb,superscriptaddress]{revtex4}

\usepackage{graphicx,epsfig}% Include figure files
\usepackage{dcolumn}% Align table columns on decimal point
\usepackage{bm}% bold math
\usepackage{times}
\usepackage{url}
\newtheorem{assumption}{Conjecture}
\usepackage{amsmath}
\usepackage{amssymb}

\begin{document}

\title{New power law signature of media exposure in human response
  waiting time distributions}

\author{Riley Crane}\thanks{To whom correspondence should be addressed.
  E-mail: rcrane@ethz.ch.}  \affiliation{Department of Management,
  Technology, and Economics, ETH Zurich, Switzerland} 
  \affiliation{Human Dynamics, MIT Media Lab, Cambridge, MA, USA}

\author{Frank Schweitzer} \affiliation{Department of Management,
  Technology, and Economics, ETH Zurich, Switzerland}

\author{Didier Sornette} \affiliation{Department of Management,
  Technology, and Economics, ETH Zurich, Switzerland} \affiliation{Swiss
  Finance Institute, c/o University of Geneva, 40 blvd. Du Pont d'Arve CH
  1211 Geneva 4, Switzerland}

\date{\today}

\begin{abstract}
  We study the humanitarian response to the destruction brought by the
  tsunami generated by the Sumatra earthquake of December 26, 2004, as
  measured by donations, and find that it decays in time as a power law
  $\sim 1/t^\alpha$ with $\alpha=2.5 \pm 0.1$.  This behavior is
  suggested to be the rare outcome of a priority queuing process in which
  individuals execute tasks at a rate slightly faster than the rate at
  which new tasks arise.  We believe this to be the first empirical
  evidence documenting this recently predicted [Grinstein, G. and R. Linsker,
(2008) Phys. Rev. E 77, 012101] regime,
  and provide additional independent evidence that suggests this ``highly
  attentive regime'' arises as a result of the intense focus placed on
  this donation ``task'' by the media.
\end{abstract}

\pacs{89.75.-k, 87.23.Ge, 89.20.-a}
\maketitle

\section{Introduction}
Who has not wondered about why things take so long to be done or to come
to fruition?  Why are we not answering correspondence, emails or
returning phone calls immediately?  When we have an idea, or develop an
intention to create something new, such as a career change, a new
relationship, or a healthier lifestyle, why does it take so long to take
action and/or see results?

Recent studies of social systems suggest a simple answer: the
distribution $P(t)$ of waiting times between cause and action performed
by humans is found to be a power law $P(t) \sim 1/t^\alpha$ with an
exponent $\alpha$ less than $2$, so that the mathematical expectation of
the waiting time ($\tau$) between consecutive events is infinite. This
power law behavior applies to the waiting time until an email message is
answered \cite{Eck}, to the time intervals between consecutive e-mails
sent by a single user and time delays for e-mail replies
\cite{Barabasi_Nature05}, to the waiting time between receipt and
response in the correspondence of Darwin and of Einstein
\cite{Oliveira_Bara}, and to the waiting times associated with other
human activity patterns which extend to web browsing, library visits and
stock trading \cite{Vasquez_et_al_06}.

These observations can be rationalized by priority queuing models that
describe how the flow of tasks falling on (and/or self-created by) humans
are executed using priority ranking
\cite{Barabasi_Nature05,Oliveira_Bara,Vasquez_et_al_06}.  Let $\lambda$
denote the average rate of task arrivals and $\mu$ the average rate $\mu$
for executing them. We can then distinguish between two different
regimes: (i) $\mu\leq \lambda$, i.e. an `overburdened' regime where tasks
arrive faster than can be executed, and (ii) $\mu > \lambda$, i.e. tasks
receive attention because they arrive with a rate below the execution
rate, which we call the `highly attentive regime' here.  Using a standard
stochastic queuing model wherein tasks are selected for execution on the
basis of random continuous priority values, Grinstein and Linsker
\cite{Grinstein1} derived the exact overall probability per unit time,
$P(t)$, that a given task sits in the queue for a time $t$ before being
executed for each of these regimes:
\begin{eqnarray}
 P(t) &\sim& \frac{1}{t^{3/2}}~,~~~{\rm for}~ \mu \leq \lambda 
\label{eq:powerlawexptheorytroisdemis} \\
P(t) &\sim& \frac{1}{t^{5/2}}~e^{-\left(\sqrt{\mu} -
    \sqrt{\lambda}\right)^2~ t}~,~~~{\rm for}~ \mu > \lambda 
\label{eq:powerlawexptheory}
\end{eqnarray}
The first regime with a value of the exponent $\alpha=3/2$ is compatible
with previously reported numerical simulations
\cite{Barabasi_Nature05,Oliveira_Bara,Vasquez_et_al_06} and with most of
the empirical data. The second regime, however, is characterized by an
exponent $\alpha=5/2$, i.e. it results in much shorter tails of $P(t)$ as
compared to the first one and does not seem to have been documented
empirically.

Hence, in this paper we provide what we believe to be the first empirical
evidence documenting the second regime by analysing a dataset of
donations described together with the methodology in Section
\ref{sec:methodology}. In Section \ref{sec:donation}, we conjecture that
the second regime may not have been previously documented because of the
effect of intense \textit{media focus} on modifying the rate of execution
of this ``donation task''.  In section \ref{sec:disc-donations}, we scrutinize
the assumptions underlying our conclusions and form an hypothesis
for the ``highly attentive regime'', that is tested in 
section \ref{sec:YouTube}. Specifically, we ask whether
or not data exhibiting an exponent of $5/2$ can be used as a signature to
identify tasks receiving significant media attention.  For this, we
investigate a massive, unique dataset containing the timeseries of the
daily view counts for nearly five million videos on YouTube.  We find
that this approach, while performed \textit{ex post}, provides compelling
results that support our \textit{media focus} conjecture.  Finally, in
section \ref{sec:discussion}, we discuss these results and propose a set
of experiments to place these results on a more firm methodological
footing.

\section{Methodology \label{sec:methodology}}

In this paper, we study the dynamics of the daily donation amount in
response to the devastating tsunami generated by the Sumatra earthquake
of December 26, 2004. This dataset was already analyzed by Schweitzer and
Mach \cite{SchweitzerMach} and it was shown that the total fraction of
donors over time, $f(t)$, can be perfectly described by an epidemic
model, $f(t)=[1+\exp\{-(t-\omega)/\tau\}]^{-1}$, provided that one takes
into account a slowing down of the inverse time scale $1/\tau(t)$. The latter has
a clear relation to mass media communication: people simply lost their
interest in the topic because of an overuse in the media (a kind of
'mental' saturation). The second parameter $\omega$ gives the time when
the donations are peaked.  As reported in Ref.~\cite{SchweitzerMach},
this value varies for different donor organizations (8 days after the
tsunami for the largest donor organization (DH) versus 36 days for the
smallest one (AH)).  These differences arise because of the relative size and scale and target demographics of these organizations --- DH is a national alliance whereas AH is more local.  
Additionally, DH possess the infrastructure for advertising and processing
large volumes of donations and, importantly, maintained a mass media presence throughout this crisis.
In the following, we only concentrate on the largest data set (DH) because of
the statistical significance.

Different from \cite{SchweitzerMach}, we do not focus on $f(t)$ but on
the growth rate $D(t)$ which gives the daily number, $D_{N}(t)$, or daily
amount, $D_{A}(t)$, of donations (see also Figs. \ref{fig:nbdonation},
\ref{fig:figure1}), which behave very similar as already noticed in
\cite{SchweitzerMach}. As a second difference to the previous
investigations, we only focus on the \emph{long-term behavior} of $D(t)$,
i.e. for sufficiently large $t$ (see also Section
\ref{sec:disc-donations}).

In order to analyze the data, we adopt an \textit{ergodic} approach which
assumes that sampling the collective responses of many individuals in
time is equivalent to sampling many realizations of the same stochastic
process.  Therefore, in order to measure the static distribution of
waiting times describing a single individual, we instead measure the
ensemble response of a large number of ``independent, identically
distributed'' individuals, each of whom is presented with the same task
simultaneously.  This approach amounts to mapping the static distribution
of waiting times describing individuals into the time domain.  This type
of mapping is only possible because of the ``singular'' nature of the
tsunami which provides a shock that allows all possible donors to become
aware of the donation ``task'' at the same distinct time.

In order to see how the distribution of rates of such responses can be
mapped to the overall probability $P(t)$ per unit time that a given task
sits in the queue for a time $t$ before being executed, consider the
tsunami that occurred on December 26, 2004. A donation associated with
this event can be considered as a task that was triggered (but not
necessarily executed) on that day simultaneously for a large population
of potential donors. This task competes with many others associated with
the job, private life and other activities of each individual in the
entire population. The specificity of the social experiment provided by
the tsunami is that the same ``singular task'' is presented at
approximately the same time to all potential donors, but the priority
value of this singular task can be expected to be widely distributed
among different individuals.  Grinstein and Linsker \cite{Grinstein2}
showed that the distribution (\ref{eq:powerlawexptheory}) is independent
of the specific shape of the distribution of priority values among
individuals. Since the singular task has been initiated at nearly the
same time for all individuals, the donations at a time $t$ after this
initiation time is then simply equal to $D(t) = N \times P(t)$, where $N$
is the number of individuals who will eventually donate in the population
and $P(t)$ is the previously defined overall probability per unit time
that a given task sits in the queue for a time $t$ before being executed.
This suggests that the tsunami donation data provides a direct test for
the prediction of the priority queuing model solved exactly by Grinstein
and Linsker \cite{Grinstein1}, and this methodology forms the basis for
the analysis we perform.

\section{Empirical Results for Donations \label{sec:donation}}

\begin{figure}[hhh]
 \centerline{\epsfig{figure=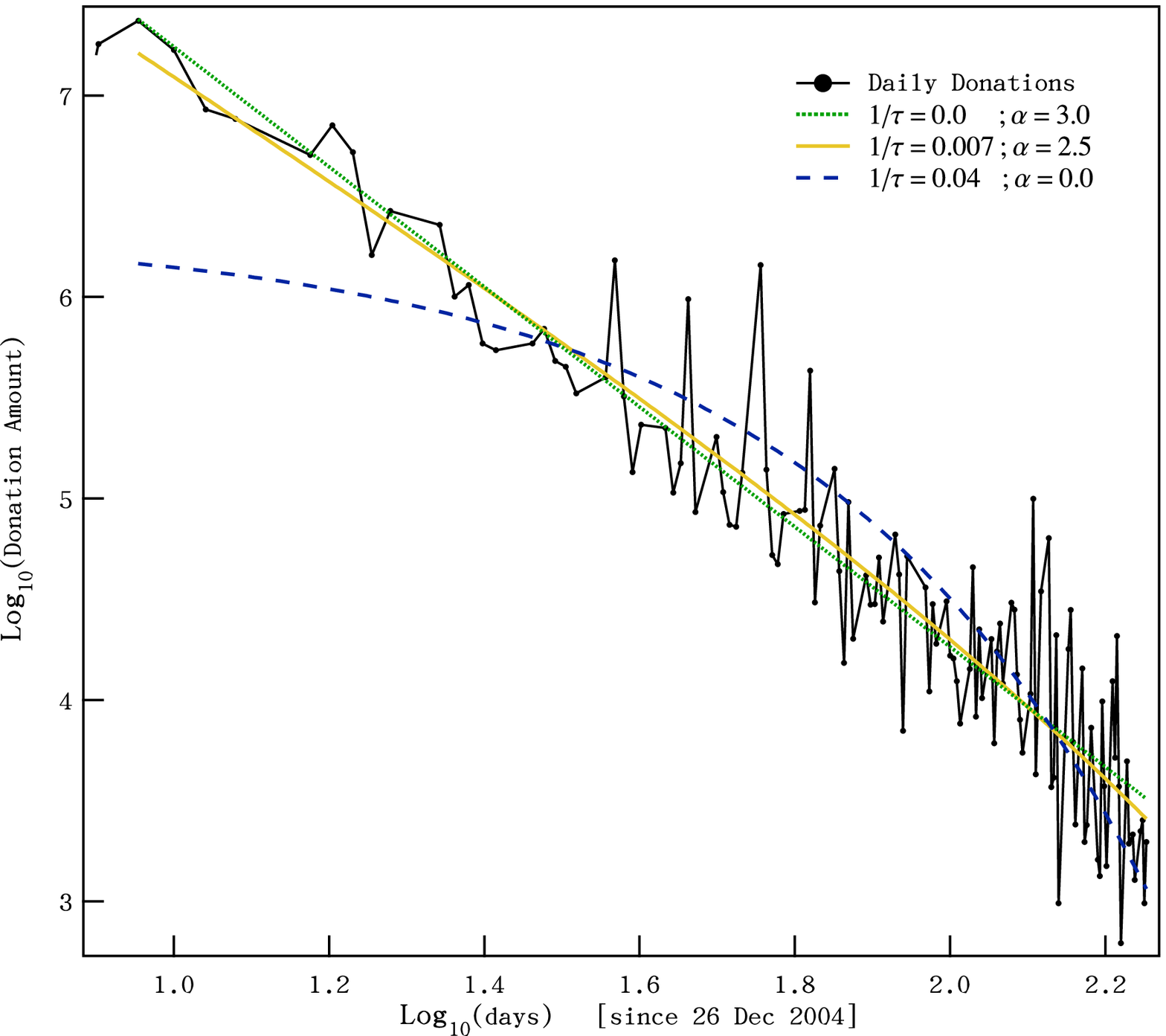,angle=0,width=11.0cm}}
 \caption{(Color online) Daily number of donations following the Asian tsunami of December 26, 2004 and its
 fits with models (\ref{eq:powerlaw}) (green line and $p$-value $=0.16$)
 and (\ref{eq:powerlawexp}) (yellow line
 and $p$-value $=0.98$). 
The pure exponential
 relaxation (corresponding to $\alpha=0$ in
 (\protect\ref{eq:powerlawexp})) is shown as the dashed black line
 ($p$-value $=0$).
 The description and source of the data is found in \protect\cite{SchweitzerMach}.
 \label{fig:nbdonation}}
\end{figure}

\begin{figure}[hhh]
  \centerline{\epsfig{figure=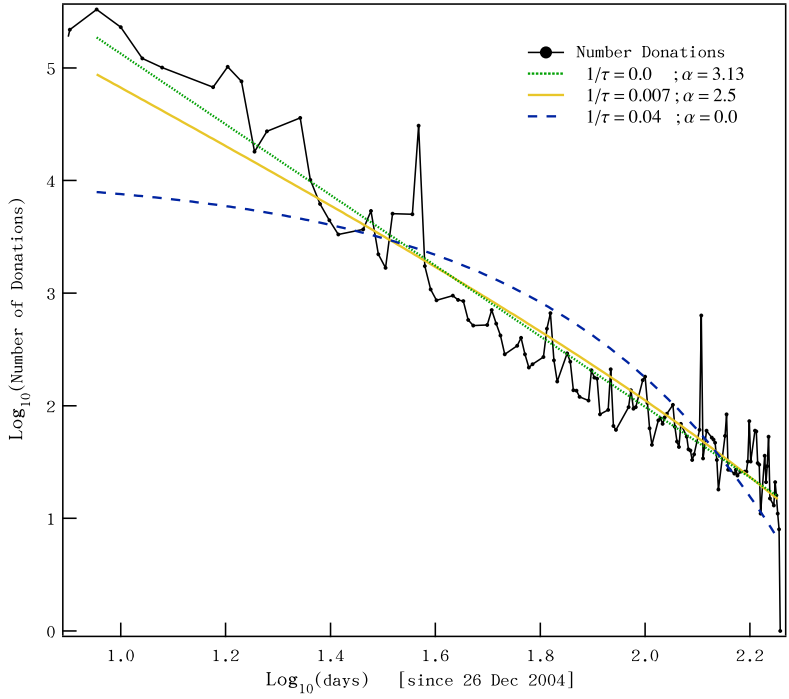,angle=0,width=11.0cm}}
 \caption{(Color online) Daily donation amounts corresponding to the data of Figure
   \protect\ref{fig:nbdonation} and its 
 fits with models (\protect\ref{eq:powerlaw}) (green line) and (\protect\ref{eq:powerlawexp}) (yellow line). 
 The pure exponential 
 relaxation (corresponding to $\alpha=0$ in (\ref{eq:powerlawexp})) is
 shown as the dashed black line.
 The description and source of the data is found in \cite{SchweitzerMach}.
 \label{fig:figure1}}
\end{figure}

Here we study both the daily number of donations, $D_{N}(t)$
(Fig. \ref{fig:nbdonation}), and the daily amount of donations,
$D_{A}(t)$ (Fig. \ref{fig:figure1}), as a function of time, where $t$
is counted in days after the tsunami. Both curves are characterized by a
large burst of donation activity in the days after the tsunami on
December 26, 2004, followed by a slow decay.  The curves peak around
January 4th, a 9 day delay likely caused by the timing of the event
around the holidays when many people were away, and charities were not
open. Schweitzer and Mach \cite{SchweitzerMach} demonstrated that the
decay in the rate of donations is slower-than-exponential, with an
instantaneous decay rate $\tau$ growing proportional to the time $t$
elapsed since the occurrence of the tsunami. This observation is typical
of scale-free dynamics and thus suggests a power law decay of the type
\begin{equation}
 D(t) = \frac{A}{t^{\alpha}}, ~~~{\rm for~large}~t~.
\label{eq:powerlaw}
\end{equation}
% $t$ is the time counted since the occurrence of the tsunami, or close to
% it to take into account finite-size effects.  
The green line in Figure \ref{fig:nbdonation} is the power law
(\ref{eq:powerlaw}) with adjustable exponent where the parameter
$\alpha=3.1 \pm 0.1$ is determined by maximum likelihood (ML).  The black
dashed line in Figure \ref{fig:nbdonation} corresponds to a pure
exponential relaxation and is clearly an inferior fit
($p$-value$=0$). The yellow line in Figure \ref{fig:nbdonation} is a fit
using
\begin{equation}
 D(t) = \frac{A}{t^{\alpha}}~e^{-\frac{t}{\tau}}~,~~~~~{\rm with~fixed}~\alpha=2.5~,
\label{eq:powerlawexp}
\end{equation}
which has the same explanatory power as the pure power law model
(\ref{eq:powerlaw}) ($p$-value$=0.16$). Allowing both $\alpha$ and $\tau$
to be freely fitted in (\ref{eq:powerlawexp}) (two-parameter model) is
undistinguishable from the one-parameter model (\ref{eq:powerlawexp})
with fixed $\alpha=2.5$ ($p$-value$=0.98$).

Figure \ref{fig:figure1} shows the total amount $D_{A}(t)$ of individual
donations for each day, corresponding to the data of Figure
\ref{fig:nbdonation}. The calibration of models (\ref{eq:powerlaw}) and
(\ref{eq:powerlawexp}) give practically the same values for $\alpha$ and
$\tau$ in all cases. This is expected from the fact that, while the
amount of a donation can vary in size, the statistical distribution of
donation sizes for this particular data has been shown to be constant
both before and after the tsunami \cite{SchweitzerMach}.  

Thus, from Figs. \ref{fig:nbdonation} and \ref{fig:figure1}, we conclude that
total the donation value per day or daily number of donations give
practically the same tapered power law decay (\ref{eq:powerlawexp}),
which holds for sufficiently large $t$.  One possible explanation for
these results is based on the priority queuing model in the regime $\mu >
\lambda$ (rate of performing tasks faster than rate of task arrival),
which predicts the exponent $\alpha=2.5$.  Interpreted in the context of
the queuing model, the donation rates following the tsunami is influenced
by the extreme media attention, which lead the potential donors to
address this task more diligently than they would otherwise.
Quantitatively, a comparison of (\ref{eq:powerlawexptheory}) and
(\ref{eq:powerlawexp}) provides an estimation of $(\sqrt{\mu} -
\sqrt{\lambda})^2= 1/\tau \simeq 0.0075 \pm 0.0001$ days$^{-1}$, leading
to $\tau \approx 130$ days and $\mu=\lambda (1+\epsilon)$ with $\epsilon
\simeq \frac{0.17}{\sqrt{\lambda}}$. Taking for instance a rate of
$\lambda=10$ tasks arriving per day, the average rate of executing them
is then estimated as $\mu \approx 10.6$ tasks per day. The closeness of
$\mu$ to $\lambda$ signals the proximity to the bifurcation from the law
(\ref{eq:powerlawexptheory}) to the much slower decay $P(t) \sim
\frac{1}{t^{3/2}}$ occurring for $\mu \leq \lambda$ \cite{Grinstein1}.
We note again that all previously mentioned examples
\cite{SornetteJohansenOriginalInternaut,SornetteAmazonPRL, Deschatres,
  cranesoryoutube, Dezso_et_al_06,Leskovec_et_al_07} report an exponent
$\alpha = 1.5 \pm 0.2$, in agreement with the queuing regime $\mu \leq
\lambda$ where the execution rate is smaller that the task arrival rate,
whereas this example clearly fits into the regime $\mu > \lambda$.  Our
present model seems quite parsimonious with just one exponent of the
power law which is predicted without adjustment to be equal to $5/2$,
plus a weak exponential taper.

\section{Discussion of Long Term Behavior}
\label{sec:disc-donations}

In this section, we want to take a closer look into the underlying
assumptions of our conclusions. Many other studies have documented a
power law decay in time of the rate of responses to shocks in human
societies which, according to the above reasoning, can be rationalized as
revealing the distribution of individual waiting time distributions. This
is precisely correct when there is no significant ``epidemic'' process in
which individuals' actions may be triggered indirectly by the initial
shock via the influence of previously active individuals.  On the other
hand, Schweitzer and Mach \cite{SchweitzerMach} have indeed reported
about the epidemics in donation behavior. In order to verify that there
is no contradiction involved, one has to consider the time scales
explicitely. As shown in \cite{SchweitzerMach}, the epidemic dynamics is
effective mostly for early times, i.e. close to the maximum of daily
donations ($\log_{10}t \approx 1$) and disappeared already at about $\log_{10} t \approx 1.4$
(January 20). The current analysis, on the other hand, holds only for the
long term dynamics and provides a reasonable fit only for times 
$\log_{10} t > 1.4$, where epidemic influences are indeed negligible, in
agreement with the derivation of (\ref{eq:powerlawexp}).

The precise mapping between the power law decay of the rate of activity
to the power law distribution of waiting times is thus valid only in the
sub-critical regime characterized by a branching ratio less than $1$
\cite{HSbasic02,SorHelm2003,cranesoryoutube}.  In the critical regime
where cascades of triggering occur, the power law decay in time of the
rate of activity is renormalized by the social epidemic process
\cite{HSbasic02}.  The precise mapping or its renormalized version apply,
for instance, in the relaxation of the rate of downloads after the
publication of an interview \cite{SornetteJohansenOriginalInternaut}, in
the relaxation dynamics of book sales on Amazon.com and video views on
YouTube$^{\rm TM}$ \cite{SornetteAmazonPRL,Deschatres,cranesoryoutube},
in the dynamics of visitations of a major news portal
\cite{Dezso_et_al_06}, and in the decay of popularity of internet blogs
posts \cite{Leskovec_et_al_07}, as well as in the relaxation of financial
volatility after a peak \cite{VolMRW}.

As a second point, we'd like to give an interpretation of the behavior
observed after January 20, i.e. in the tail of the relaxation of the
donation rate after the main peak of activity. In accordance with the
theoretical prediction \cite{Grinstein1}, we observed a ``highly
attentive regime'' characterized by $\mu > \lambda$, i.e. the effective
rate of performing tasks is larger than the rate of arriving tasks.
Different from the early times (at about the maximum of donations per
day) where herding played the dominant role, people donating at later
times were really in the regime to act on purpose (``attention''), rather
than being ``forced'' to donate. 

We argue that the ``highly attentive regime'' is caused by an effective
increase in $\mu$, rather than by an effective decrease in $\lambda$, and
that the increase in $\mu$ is the result of an extensive media coverage
after the tsunami desaster.  By being continuously exposed to the
devastating effects of the tsunami, individuals which have not donated so
far have seen their execution donation rate renormalized upwards into
what could be called the ``highly attentive'' regime $\mu > \lambda$,
from what appears to be the more ubiquitous ``overburdened'' regime $\mu
\leq \lambda$.

Rather than increasing the effective rate $\mu$ of performing tasks, one
could argue that a large part of the donations were not triggered by the
tsunami itself, but by the intense media coverage that extended
continuously over the period covered by our data. This would then predict
a broad distribution of ``task arrival times'' spanning the whole time
interval over which the media was highly active.  But, together with the
usual ``overburdened'' regime $\mu \leq \lambda$ leading to a slower
relaxation of the donation rate with exponent $\alpha =3/2$, this would
create an even slower decay of the relaxation rate, given roughly by the
convolution of the broad distribution of media-triggered task arrival
times with the slow response to each of them. We conclude that this
cannot be the explanation for the much faster decay
(\ref{eq:powerlawexp}).  The impact of the media attention is most likely
to result in the creation of a ``highly attentive regime,'' leading to an
increased focus placed on this donation ``task'', quantified by $\mu >
\lambda$.  The next step is to validate our hypothesis on a different
data set, which we proceed to perform in the next section.

\section{Preliminary validation using YouTube data \label{sec:YouTube}}

To our knowledge, the tsunami donation data provides the first empirical
example of the second predicted regime $\mu > \lambda$ \footnote{After
  this paper was made available online, we were informed by the authors
  of a paper on bloggers behavior that they also found an exponent larger
  than 2 in their data. M Mitrovic, B. Tadic: Bloggers Behavior and
  Emergent Communities in Blog Space, European Physical Journal B
  (2010)}. Compared with the exponents of the power law rate of responses
to shocks in human societies discussed above which are all of the order
of or smaller than $3/2$, the tsunami donation experiment is arguably the
only one with such exceptional media coverage, both in intensity and
duration.  We conjecture that the intense pressure by the combined media
led to an effective increase of the rate of execution of this donation
``task.'' Our conjecture can be formulated more generally as follows.

\begin{assumption}
\label{conj1}
\item
\begin{enumerate}
\item If a social activity following a singular task presented at
  approximately the same time to a population enjoys continuous media
  exposure or is subjected to other persistent attention grasping
  processes, we should expect it to decay according to
  (\ref{eq:powerlawexp}) (as predicted by (\ref{eq:powerlawexptheory}))
  if social interactions can be neglected (corresponding to the exogenous
  sub-critical regime, in the classification of
  Ref. \cite{cranesoryoutube}).

\item Reciprocally, if an activity following some shock is observed to
  decay according to (\ref{eq:powerlawexp}), one should expect to find
  evidence for continuous media exposure or for other mechanisms
  promoting the ``highly attentive'' regime.
\end{enumerate}
\end{assumption}

In the following, we discuss a preliminary verification of the second
element of conjecture \ref{conj1}.  By analyzing the relaxation response
of the daily view counts for nearly 5 million videos on the popular video
sharing website YouTube, we performed an \textit{ex post} investigation
to determine whether data with the characteristic relaxation signature of
$1/t^{5/2}$ could be clearly linked to events which received significant
media focus.  Briefly, each video is described by a timeseries of its
daily view count, collected each day over a period of up to 1.5 years.
We extract the relaxation exponents using a least-squares fit on the
logarithm of the data.  A full description of this data and the
methodology used for analysis has been previously reported in
\cite{cranesoryoutube}.  Selecting timeseries whose daily view count
experiences a burst followed by a relaxation best fit given by
\begin{equation}
 V(t) \sim \frac{1}{t^{\alpha}}~,
\label{eq:youtube}
\end{equation}
where $V(t)$ is the daily view count and $\alpha = 2.5 \pm 0.1$, returns
$7,230$ matching videos.

\begin{figure}[hhh]
 \centerline{\epsfig{figure=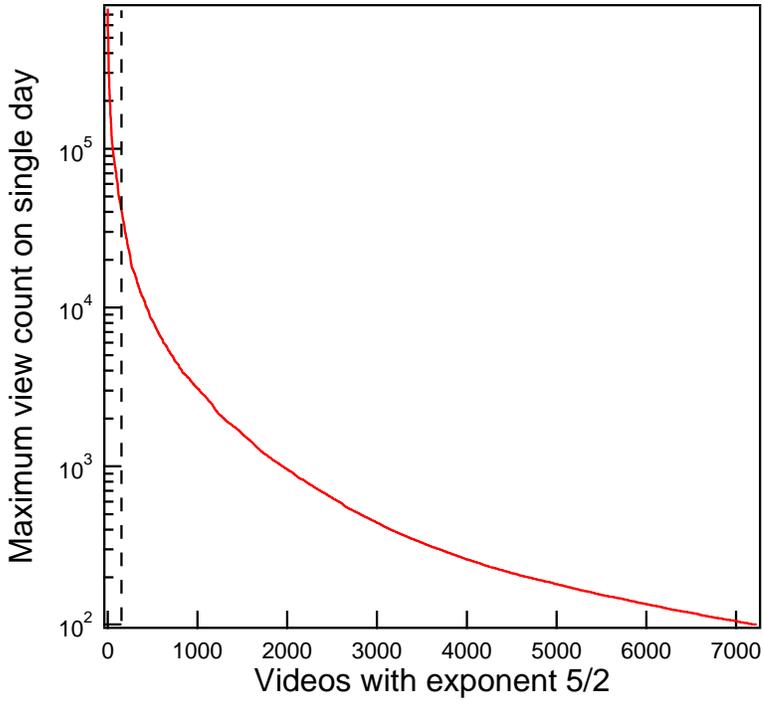,angle=0,width=11.0cm}}
\caption{(Color online) Maximum number of views obtained on a single day for each of the
$7,230$ videos which experience a burst followed by a relaxation of
the form $V(t) \sim \frac{1}{t^{\alpha}}$ with $2.4 \le \alpha \le
2.6$.  The dashed vertical line separates the $124$ videos which
received more than $50,000$ views on a single day from those which
received less.  \label{fig:max_view_count}}
\end{figure}

Figure \ref{fig:max_view_count} shows the $7,230$ videos which are ranked
according to the maximum view count received on a single day.  As a
condition for the application of the conjecture, we should only consider
the videos which experienced a large, sudden ``burst'' of activity, since
this ensures that social interactions can be neglected and the relaxation
dynamics falls into the exogenous sub-critical regime according to the
classification of Ref. \cite{cranesoryoutube}. We thus focus our
attention on the top 124 videos that each received more than $50,000$
views on a single day, as delineated with the dashed verticle line in
Figure \ref{fig:max_view_count}.

\begin{figure}[]
\centerline{\epsfig{figure=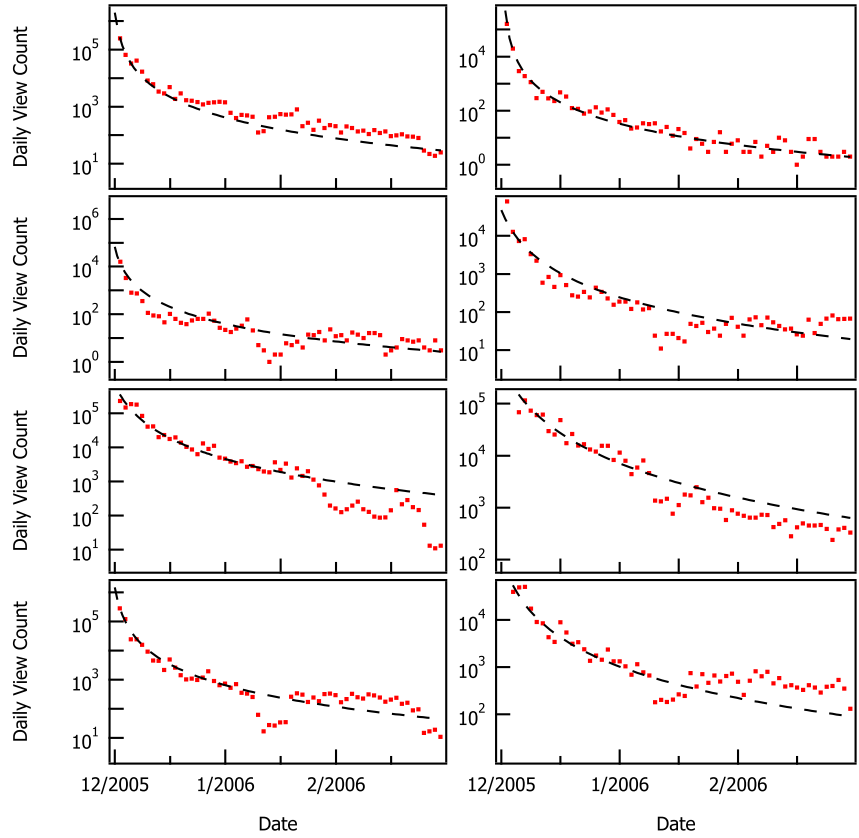,angle=0,width=12.0cm}}
\caption{(Color online) Eight log-lin plots showing the dynamics of the videos related to the execution of Saddam Hussein.  Fits to each dataset are shown as a dashed line and described by a power law relaxation with an exponent in the range $2.4 \le \alpha \le 2.6$.  \label{fig:saddam}}
\end{figure}

While it is not possible to detail the context and content of each of the
124 videos, some very clear trends emerge.  Examining these videos
carefully, we find that 37 ($\approx 30\%$) refer to items for which we
could find news articles or press releases.  Of these, 24 ($\approx
19\%$) refer directly to news events (political figures, celebrities,
etc), 5 ($\approx 4\%$) are related to movie releases, and the most
interesting result is that 8 ($\approx 7\%$) are on the topic of Saddam
Hussein's execution.  The dynamics for these last eight videos are shown
in figure \ref{fig:saddam}, and we see they are very similar indeed to
the donation data.

While these results certainly do not prove our conjecture that media
focus can modulate our innate task execution rate $\mu$, they support our
interpretation and provide \textit{plausible} evidence that such a
mechanism could be responsible for the observed exponent value of $5/2$
in the empirical data.  This conclusion is further supported by the fact
that an exhaustive analysis of our entire YouTube database reveals that
the typical decays of views are characterized by exponents clustering
around three values $\alpha=0.2$, $\alpha=0.6$ and $\alpha=1.4$, which
have been understood as respectively associated with endogenous critical,
exogenous critical and exogenous subcritical regimes
\cite{cranesoryoutube}. The value $\alpha \approx 2.5$ found here is thus
very atypical and its observation associated with particular ``highly
attentive'' regimes is particularly revealing.

\section{Conclusion and Outlook \label{sec:discussion}}

The contributions of the present work are (i) the empirical support for
the newly proposed regime describing individual waiting time
distributions and (ii) the further support provided to the interpretation
that this regime results from a media enhanced task execution rate that
modifies the distribution with exponent $3/2$ to a thinner tailed
distribution with larger exponent $5/2$.  These results open a number of
possibilities to quantitative studies on the manipulation of execution
rates of ensembles of individuals.  A more rigorous test of this approach
would search for ``singular events'' which propose some ``task'' to be
completed, and determine whether one can obtain some quantitative measure
of attention grasping mechanisms in the form of search queries, video
views, purchases, downloads, or some other suitably well-defined
quantity.  While we believe we have found strong evidence supporting our
claim to have observed a new regime governing individual dynamics,
designing an experiment to test the conjecture \textit{a priori} would
provide direct evidence.

Finally, this work emphasizes the important contributions that can be
made to understanding individual waiting time distributions by
investigating the dynamical collective response to ``shocks'' or other
events with a clearly identifiable footprint in time.  While other
studies have implicitly used this approach to rationalize their data, we
do not believe this point has been stressed, or that this approach has
been fully exploited in the literature.  It may not be per se surprising
that media focus can enhance our implicit rate for taking action.
However, it is quite remarkable that this can produce clearly
identifiable signatures in the dynamics of collective action. Hence,
these findings may open a new frontier for the quantitative study of
individual and social activity.

Further work would involve combination
of  similar data for other comparable activities, such as for the Haiti earthquake
in January 12, 2010,
but combined with social studies, such as questionnaires,
interviews and other tools probing the real motivation and media
exposure of donors. The Haiti disaster could also be a good testing
case, as much of the media coverage consisted of descriptions of
mishandling the international help, and such coverage could, 
according to our reasoning, significantly decrease the donations.

% That media focus can enhance our implicit rate for taking action is not
% all that surprising, however that this can produce clearly identifiable
% signatures in the dynamics of collective action opens a new frontier
% for the quantitative study of individual and social activity.

\clearpage

\clearpage

\clearpage

\clearpage

\end{document}